\documentclass{JHEP3}
\usepackage{epsfig,amsmath,amsthm}
\newcommand{\be}{\begin{equation}}
\newcommand{\ee}{\end{equation}}
\newcommand{\bea}{\begin{eqnarray}}
\newcommand{\eea}{\end{eqnarray}}
\def\bse{\begin{subequations}}
\def\ese{\end{subequations}}
\newcommand{\IR}{\mathbb{R}} 

\def\IZ{\relax\ifmmode\hbox{Z\kern-.4em Z}\else{Z\kern-.4em Z}\fi}
\newcommand{\IS}{\mathbb{S}}

\newcommand{\non}{\nonumber \\}

\def\half{{1 \over 2}} 

\def\del{{\partial}} \def\dag{\dagger}

\def\cl{{\cal L}} \def\co{{\cal O}}
  
\def\cL{{\cal L}} 

\def\tphi{\widetilde{\phi}}

\def\al{\alpha} \def\bt{\beta}

\def\presub{\vspace{.5cm} \noindent}

\def\bi{\begin{itemize}} \def\ei{\end{itemize}}

\def\Schw{Schwarzschild }
\def\({\left(} \def\){\right)}
\def\[{\left[} \def\]{\right]}
\preprint{{\tt hep-th/0609001}}

\title{ \center{Perturbations Around Backgrounds\\ with One Non-Homogeneous Dimension}}

\author{
Barak Kol \\
 Racah Institute of Physics\\
 Hebrew University \\
 Jerusalem 91904,
 Israel\\
{\tt barak\_kol@phys.huji.ac.il}}

\abstract{This paper describes and proves a canonical procedure to
decouple perturbations and optimize their gauge around backgrounds
with one non-homogeneous dimension, namely of co-homogeneity 1,
while preserving locality in this dimension. Derivatively-gauged
fields are shown to have a purely algebraic action; they can be
decoupled from the other fields through gauge-invariant
re-definitions; a potential for the other fields is generated in the
process; in the remaining action each gauge function eliminates one
field without residual gauge. The procedure applies to spherically
symmetric
 and to cosmological backgrounds in either General Relativity or gauge
theories.
 The widely used ``gauge invariant
perturbation theory'' is closely related. The supplied general
proof elucidates the algebraic mechanism behind it as well as the
method's domain of validity and its assumptions.}

\begin{document}



\section{Introduction}

Consider perturbations around backgrounds with one non-homogeneous
dimension (namely, with co-homogeneity one). This category
encompasses two important and widely studied classes: spherically
symmetric, static backgrounds where the radial coordinate $r$ is the
essential, non-homogeneous coordinate, and homogeneous cosmologies
where time is the essential coordinate.

Technically, the problem may be reduced through separation of
variables, expansion in terms of harmonic functions over the
homogeneous spaces and dimensional reduction, to a 1d action
invariant under certain gauge transformations. The issues are to
optimally decouple the fields while choosing an appropriate gauge.

Recently, a specific case which belongs to this class was solved,
that of the negative mode around the \Schw black hole background
\cite{nGPY}. The analysis followed by postponing any gauge-fixing;
obtaining the action in terms of a ``maximally general ansatz'' (one
which takes account of isometries yet is general enough to produce
all equations from the variation of the action); using the power of
the action formalism to transform the fields and decouple them into
two sectors, algebraic and dynamic; and finally solving the gauge
away, altogether reducing a problem of 5 fields and 2 gauge
functions into a master equation for a single dynamic field. It was
suggested there that the procedure could be generalized.

In this paper we make good that suggestion of \cite{nGPY}: we
formulate the generalization in the form of a theorem and supply the
proof, along with its conditions and limitations. We start in
section \ref{theorem-section} by setting up the problem and
notation, and distilling the data for the general case. Then we
choose a path to tread which leads us to the characterization of the
algebraic fields in terms of {\it derivatively gauged} fields which
we define, and to the formulation of theorem \ref{thm}, which is our
central result. After stating the theorem we prove it.

Our procedure overlaps with ``gauge invariant perturbation theory''
which is widely used in the literature. In section
\ref{discssion-section} we discuss its added value, and we provide
further detail on the application to spherical symmetry and to
cosmological backgrounds.

\section{Theorem}
\label{theorem-section}

{\bf Set-up}. Consider a 1d quadratic action with a gauge symmetry.
We denote the 1d variable by $x$, and the fields and gauge fields
respectively by the vector notation \bea
 \phi&=&\phi^i(x) \qquad i=1,\dots,n_F \non
 \xi&=&\xi^a(x) \qquad a=1,\dots,n_G \eea
where $n_F,\, n_G$ are the number of (real) fields and gauge
functions, respectively.

We wish to generalize the procedure of \cite{nGPY} to decouple the
action. Since at this point it is not clear how to do that, we start
by \emph{distilling the data} for the general case.

The most general action quadratic in fields which is at most
quadratic in derivatives is  \bea
 S &=& S[\phi,\phi]= \int dx\, \cL[\phi,\phi] \non
 \cL[\phi,\phi] &:=&  \phi'\,^T\, L_2(x)\, \phi'\,  +  \phi^T\, L_1\, \phi' +
 \phi^T\, L_0\, \phi   \non
 &:=&  L_{2,ij}(x)\, \phi^i\,'\, \phi^j\,' +  L_{1,ij}(x)\, \phi^i\, \phi^j\,' +
 L_{0,ij}(x)\, \phi^i\, \phi^j    \eea
where $S[\phi,\phi]$ is a condensed operator notation for a
quadratic form which operates on $\phi^i(x)$ which the second line
describes in vector notation in terms of $\(L_2(x),\, L_1(x),\,
L_0(x) \)$, a triplet of $n_F \times n_F$ $x$-dependent matrices and
finally the third line spells out the vector notation. $L_0(x)$ and
$L_2(x)$ are symmetric, while since the action is defined up to the
addition of $\del_x \( \phi^T\, \hat{L}\, \phi \)$ where $\hat{L}$
is any symmetric matrix, $L_1(x)$ can be assumed anti-symmetric.

The gauging is linear in the gauge functions, independent of fields
and we assume it is at most linear in derivatives, which is indeed
the case when analyzing perturbations in GR or gauge theory. It is
given by \be
 \delta \phi = G\, \xi := G_1\, \xi' + G_0\, \xi \ee
where the condensed operator notation of the first equality is
exchanged by a vector notation in terms of $\(G_1(x),\,G_0(x) \)$, a
pair $n_F \times n_G$ $x$-dependent matrices.

Gauge invariance ties $\cL$ and $G$ by requiring \bse \label{LG} \be
 0= \half\, \delta \cL = \cL[\phi,\, G\, \xi]  \ee
 for all $\phi^i,\, \xi^a$,  and hence in condensed notation, which will be expanded later \be
 \cL\, G =0  ~.  \label{LGb} \ee \ese

Altogether the quadratic form $S[\phi,\phi]$ holds all the data,
including that of the gauging which must be (a subset of) its
symmetries. Our purpose is to decouple parts of the action as much
as possible. In other words, we wish to ``diagonalize'' $S$, to
bring it to a canonical form. Of course $S$ (or $\cL$) is not an
ordinary matrix whose entries are numbers, but rather they are
differential operators, or equivalently polynomials in $\del_x$.
\footnote{An algebraic structure consisting of ``vectors'' whose
entries are elements of a ring, such as the polynomials, rather than
a field (such as the real or complex numbers) is called a
``module''. Accordingly, matrices with polynomial entries are linear
transformations between modules. We are especially interested in the
ring of polynomials in a single variable $\del_x$. This ring allows
for the procedure of division with remainder and is therefore said
to be an ``Euclidean ring''.}
 \newcounter{footnotemodule}
\setcounter{footnotemodule}{\value{footnote}}
 These observations lead us to an \emph{algebraic approach} to the
problem, namely to find the invariant properties of the quadratic
form $S$.

Examples such as \cite{nGPY} show that the action contains a {\it
mutually algebraic} sector, namely a subspace over which the
restriction of $\cL$ is algebraic and hence identical to the
restriction of $L_0$. How could that subspace be characterized? The
following definition is central

\theoremstyle{definition}
\newtheorem{theorem}{Theorem}
\newtheorem{definition}{Definition}

\begin{definition} \label{def}
The image of $G_1$ in field space is called {\it the
derivatively-gauged (DG) fields}.
\end{definition}

\noindent \emph{Notation}. $L_{DG}$ denotes the restriction of $\cL$
to the DG-fields.

We may now state the theorem

\begin{theorem} \label{thm}
 Given a 1d quadratic action with a gauge symmetry,\footnote{Where
the action is at most quadratic in derivatives, and the gauging is
at most linear.} the following canonical procedure can be applied to
decouple it into two sectors, algebraic and dynamic, and completely
eliminate the gauge:
\begin{enumerate}
 \renewcommand{\labelenumi}{\Alph{enumi}.}
 \item The {\it derivatively gauged} (DG) fields are {\it mutually
 algebraic}.
 \item The DG-fields decouple from the rest after a shift in
 their definition (assuming the algebraic $L_{DG}$ is invertible),
  or more precisely a certain shift in the
 definition of coordinates on field space.
 \item The shifted field coordinates are gauge-invariant.
 \item The remaining action is algebraically gauged\footnote{Namely, $G_1=0$ when restricted to this sector.}
and as such each gauge function can be used to eliminate a field
ending up with a {\it dynamic} action in $n_F-2\, n_G$
gauge-invariant fields (under the genericity assumption that both
$G_X$, to be defined below, as well as $G_1$ are non-degenerate).
\end{enumerate}
 \end{theorem}

\presub {\bf Proof}. All of the algebraic information is contained
in the explicit expansion of the gauge invariance identity
\eqref{LG} into powers of $\del_x$ (integration by parts allows to
transfer derivatives from $\phi$ to $\xi$ and vice versa so only the
total number of derivatives matters). It is instructive to assume
the special case where the $L$ matrices are $x$-independent. In this
case one may Fourier transform the action and obtain $\cL$ as a
matrix whose entries are quadratic polynomials in $k$, the Fourier
conjugate of $x$, and similarly the entries of $G$ are linear
polynomials in $k$. The gauge invariance identity \eqref{LGb} can be
expanded into powers of $k$ from $k^0$ up to $k^3$ yielding 4
$k$-independent identities (between constant matrices!), which is
all the information one has, and they turn out to exactly suffice
for the proof. Here however we shall start directly with the most
general case where the $L$ matrices do depend on $x$.

\presub \emph{Item A}. From the third (and highest) order of
\eqref{LG} in $\del_x$ we find \bse \be
 0=\half\, \delta \cL =  \phi'\,^T\, L_2\, G_1\,
\xi'' + \co\(\del_x\,^2\)  ~.\ee
 Since the equation holds for all
$\phi^i,\, \xi^a$ we deduce that \be
 L_2\, G_1 =0 \label{LG3} \ee \ese
with the interpretation that the DG-fields do not participate in
$L_2$, the kinetic part of the action.

Moreover, since $\delta S=0$ identically so does the second
variation vanish \bse \bea
 0 = \half\, \delta^2 \cL &=&  (G_1\, \xi'')^T\, L_2\, G_1\,
 \xi'' + \non
 &+& 2\, (G_1\, \xi'')^T\, L_2\, (G_0+G_1') \xi' + (G_1\, \xi')^T\, L_1\, G_1\, \xi'' +
 \non
 &+& \co\(\del_x\,^2\)  = \non
 &=&  (G_1\, \xi')^T\, L_1\, G_1\, \xi'' +  \co\(\del_x\,^2\)   \eea
 where in the last equality we used \eqref{LG3}.
 Since the equation holds for all $\xi^a$ we deduce that
\be
 G_1^T\, L_1\, G_1 =0 \label{SG2} \ee \ese
This means that once the action is restricted to the DG-fields not
only are the 2-derivative terms absent, but also so are the first
derivatives \be
 \cL[G_1\, \xi,\, G_1\, \xi] = (G_1\, \xi)^T\, L_0\, (G_1\, \xi) \ee
 and hence, by definition, the DG-fields are {\it mutually  algebraic}.

\presub \emph{Item B}. We split the field space $\Phi$ into the
DG-subspace $DG$ with coordinates $\phi_{DG}^\al$ and an arbitrarily
chosen complement $X$ with coordinates $\phi_X^r$, namely \be
 \Phi = DG \oplus X \ee or equivalently
\be
 \phi^i=(\phi_{DG}^\al,\, \phi_X^r) ~.\non \ee

The action can be cast into the following form \be
 \cL[\phi_X,\phi_X] = \phi_{DG}^T\, L_{DG}~ \phi_{DG} + 2\, \phi_{DG}^T\, (\cL_m\,
 \phi_X) + \widehat{\cL_X}[\phi_X,\phi_X] \ee
  through a step which involves integration by parts and
 yields the definition of the mixed part to be \be
 \cL_m = L_1\, \del_x + (L_0 + \half\, L_1') ~.\ee
The types of indices of $L_{DG}$ are $L_{DG} \equiv L_{DG}\,_{\al
\bt}$ and we have $L_{DG}\,_{\al \bt}=L_0\,_{\al \bt}$, for $\cL_m$
they are $\cL_m \equiv \cL_m\,_{\al r}$ and finally for the
remaining $X$-part of the action the types of indices are
$\widehat{\cL_X} \equiv \widehat{\cL_X}\,_{r s}$.

 The square may be completed by defining the {\it shifted field coordinates} \be
 \tphi_{DG} := \phi_{DG} + L_{DG}^{-1}\, \cL_m\, \phi_X \label{DG-shift} \ee
 where we used our assumption that $L_{DG}$ is
invertible to achieve a decoupling of the action into two sectors
$\cL[\tphi_{DG},\tphi_{DG}]$ (algebraic) and $\cl_X[\phi_X,\phi_X]$
(possibly dynamic) \bea
 \cL[\phi,\phi] &=& \cL_{DG}[\tphi_{DG},\tphi_{DG}] + \cL_X[\phi_X,\phi_X] \non
 \cL_{DG} &:=& L_{DG} \non
 \cL_X &:=& \widehat{\cL_X}  - \cL_m\,^\dag\, L_{DG}^{~-1}\, \cL_m
 \label{S-decoupled} \eea
where $\cL_m^\dag$ denotes the adjoint operator of $\cL_m$.
\footnote{By definition $\forall \phi_1,\phi_2~~\int dx (\cL_m\,
\phi_1)\, \phi_2 = \int dx\, \phi_1\, (\cL_m\,^\dag\,\phi_2)$ .}

Let us make two comments on the decoupling. First, the change of
variables into the shifted fields \eqref{DG-shift} augmented by the
identity on $\phi_X$, namely $\widetilde{\phi_X}^r:=\phi_X\,^r$ is
an invertible change of variables, or more precisely, it has unit
Jacobian. Second, note that $\cL_X$ receives an extra term in the
process $\Delta \cL_X = -\cL_m\,^\dag\, L_{DG}^{~-1}\, \cL_m$
\eqref{S-decoupled}.\footnote{Such a process of decoupling or
eliminating algebraic fields is commonly called ``integrating out'',
a name which originates from a path integral perspective.} $\cL_m$
is at most linear in derivatives since the $DG$ fields do not appear
in the kinetic term (see the proof for item A above). Combining that
with $L_{DG}$ being algebraic we deduce that $\cL_X$ still contains
no more than 2 derivatives.

One may wish for a more geometric interpretation of the decoupling,
especially since the choice of the $\phi_X$ fields seemingly
introduced some arbitrariness into the process. Geometrically
decoupling is tantamount to finding $DG^\perp$ -- a subspace which
is the $S$-complement to $DG$. Indeed $DG^\perp$ is given by the
equations $\tphi_{DG}^\al=0$ (while $DG$ is given by the equations
$\phi_X^{~r}=0$). Thus the arbitrary choice of the complement $X$ is
replaced by the unique $S$-complement $DG^\perp$. At first, our
intuition with vector spaces may lead us to assume that any
sub-space $U$ has an $S$-complement. However, in our case where the
components of vectors and matrices are polynomials rather than
numbers\footnotemark[\value{footnotemodule}] we should exercise
caution and indeed it is found that the existence of the complement
depends on the invertibility of the restriction of $S$ to $U$.
\footnote{Indeed, while every subspace $U$ still defines an
$S$-orthogonal subspace $U^\perp$, the two do not necessary span the
whole space. For example take the module $\IZ^2$ (over the ring of
integers) with the quadratic form $S=[ 2 ~ -1;\, -1 ~ m]$ where $m$
is any integer and take $U$ to be the subspace generated by $u=[1~
0]$. Now $U^\perp$ is generated by $u^\perp=[1~ 2]$; However $u$ and
$u^\perp$ do not span $\IZ^2$, and that can be traced back to the
fact that $S_U \equiv [2]$ is non-invertible.}

\presub \emph{Item C}. Requiring gauge invariance of the action in
its decoupled form \eqref{S-decoupled} we find that the piece which
includes $\tphi_{DG}$ is \be
 0=\half\, \delta \cL = \tphi_{DG}^T L_{DG}^{~-1}\, \delta\tphi_{DG} + \dots \ee
Since the vanishing should hold for any $\tphi_{DG}$ and since
$L_{DG}^{~-1}$ is invertible we immediately obtain  the required \be
 \delta \tphi_{DG} =0 ~.\ee
Note that in this proof we used the gauge invariance equation
\eqref{LG} at orders $0,1$ in $\del_x$, thus completing an all-order
use of that identity.

\presub \emph{Item D}. By definition $\phi_X$ are algebraically
gauged. If we denote by $G_X \equiv G_{0X}$ the gauging when
restricted to $DG^\perp$, we define the dynamic space $D$ to be the
gauge invariant space, namely the quotient \be
 D:= DG^\perp/Im(G_X) \ee
  $\cL_X$ is naturally defined on the quotient $D$ since $0=\half\, \delta \cL_X = \cL_X[G_X\, \xi,\, \phi_X]$,
and we call it \emph{the dynamic action}, to distinguish it from
$\cL_{DG}$ which is algebraic.

Let us count the dimension of $D$ \bea
 \dim D &=& \dim X - \dim \, {\rm Im} (G_X) \non
 \dim X &=& n_F - \dim \, {\rm Im} (G_1) \eea
Therefore, whenever $G_1$ and $G_X$ are non-degenerate we have $\dim
\, {\rm Im} (G_X)= \dim \, {\rm Im} (G_1)=n_G$ and $\dim D=n_F- 2\,
n_G$, as required.  \qed

\presub \emph{Relaxing the assumptions}. During parts of the proof
we made the following genericity assumptions: that $L_{DG}$ is
invertible, and that $G_1$ and $G_X$ are non-degenerate. We are not
interested here in trying to relax these assumptions, though it
could very well be that not all of them are necessary. For instance,
if $G_1$ were degenerate it would mean that one of the gauge
functions appeared only algebraically and as such it could serve to
eliminate a field and the situation would reduce back to the case of
non-degenerate $G_1$.

\section{Discussion and applications}
\label{discssion-section}


Our central result is theorem \ref{thm} regarding 1d quadratic
actions, which include perturbations around backgrounds with
essentially one dimension, such as static spherically symmetric
backgrounds or homogeneous cosmological backgrounds, in systems with
gauge -- be it either General Relativity of gauge theory. The result
contains several messages \bi
 \item The gauge can always be eliminated in
 favor of gauge-invariant variables, while keeping the action local (in
 $x$).
 \item Each gauge function ``shoots twice'' -- first, it is
 responsible for an algebraic field (``constraint''), which decouples from the dynamic action, and second it
 is used to eliminate a field.
 \item The algebraic fields are precisely the ``derivatively-gauged'' (DG) fields (see definition
 \ref{def}).
 \item The proof follows step by step the expansion of the gauge variation identity
 \eqref{LG}.
 \ei

\presub {\it Open questions}. Several interesting directions for
generalization remain open. Here we discussed the quadratic action,
but it seems plausible that higher orders in perturbation theory can
also be written in terms of the gauge-invariant fields. If we
consider actions with more than 2 derivatives there seems to be no
reason for a decoupled algebraic sector, but still there should be
some canonical form, just like the one for a matrix over polynomials
in a single variable. Finally, it is completely unclear at the
moment how to proceed in dimension -- to 2 essential dimensions and
higher.

\presub {\it ``Gauge-invariant perturbation theory''}. Our procedure
overlaps with the widely used ``gauge-invariant perturbation
theory'' -- the formulation of the equations solely in terms of
gauge invariant fields. Moncrief (1974) \cite{Moncrief1974} found
the advantages of gauge invariant variables for perturbations around
spherically symmetric background using a variational principle (much
like the action method in \cite{nGPY}). The constraints were
exhibited for general backgrounds and a detailed analysis was
carried for the \Schw background. Motivated by the analysis of
collapse of a star Gerlach and Sengupta (1978)
\cite{GerlachSengupta} applied it to perturbations around
time-dependent homogeneous backgrounds and generalized by
introducing the most general perturbed energy-momentum tensor. The
method of Bardeen (1980) \cite{Bardeen} is mathematically equivalent
is some respects but aims for cosmological applications. See
\cite{MukhanovFeldmanBrandenberger} and references therein for a
review of this method when applied to cosmological perturbations,
and see \cite{MartelPoisson,NagarRezzolla} for more recent
discussions of perturbations of the \Schw black hole in the
gauge-invariant formalism.

The author does not pretend to be much of a historian, but let us
attempt to discuss the value added by the current paper. Comments
and essential relevant references are welcome. \bi
 \item We prove the procedure in full generality thereby clarifying the mechanism
responsible for it, namely the algebraic properties of $S$, the
quadratic form which is the action, and in particular the identities
encompassed by (\ref{LG}).

In the literature it is commonly stated that the purpose of
introducing the gauge invariant variables is to get rid of gauge
ambiguity.\footnote{For example
``the formulation removes the annoyance and drudgery of dealing with
ambiguities due to infinitesimal coordinate transformations''
\cite{GerlachSengupta} (section I) as well as ``We shall remove the
necessity of referring to such solutions [gauge modes - BK] at all
by introducing a set of ....`gauge invariants' '' (section IV). ``To
have genuine physical significance gauge independent quantities
should be constructed from the variables naturally present in the
problem, here the perturbations in the metric tensor and
energy-momentum tensor'' \cite{Bardeen} (section 3). ``In this
review we develop a formalism ... [of] gauge invariant variables...
[T]he physical interpretation of the results in unambiguous.''
\cite{MukhanovFeldmanBrandenberger} (section 1).}
 Indeed gauge invariant variables are gauge invariant. However, this
argument is misleading since the non-trivial content of the
procedure is that there exists an invertible and local
transformation (local in $x$) to these variables. To the best of the
author's knowledge this fact is proven here for the first time in
generality.

\item Our proof relies on the action being 1d, or more
precisely on demanding that the transformation into gauge invariant
fields be local only in a single dimension. In 2d (and higher)
$\cL,\, G$ become polynomials in 2 variables $\del_\mu, ~\mu=1,2$;
there is no unique definition of DG-fields, and there is no general
principle to guarantee an algebraic sector. This is an important
demarkation of the domain of validity of the procedure, which again
appears here for the first time to the best of the author's
knowledge.

\item Our procedure recognizes the importance of the
``derivatively-gauged fields'' (see definition \ref{def}). \ei

\presub {\bf General properties of the action}. So far we discussed
the solution of 1d actions. Before we turn to applications let us
discuss some general properties of the construction of these actions
for perturbations around co-homogeneity 1
backgrounds.\footnote{Namely, backgrounds where all but one
coordinate belong to a homogenous space such as $\IR^3$ or $\IR
\times \IS^2$.}

{\it A priori} our procedure stresses the importance of using the
action formalism in GR which in turn requires the use of a
``maximally general ansatz'' -- an ansatz such that all the
equations of motion can be derived from varying the action. However,
with hindsight a {\it ``short-cut''} ansatz is possible: it is
enough to compute the action as a function of the fields appearing
in an ansatz with all the $n_G$ $DG$ fields as well as some $n_F-2\,
n_G$ subset of the other fields to represent the dynamic sector.
This is allowed because by gauge invariance the last $n_G$
algebraically-gauged fields do not appear in the action anyway, but
we shall not use it in this paper.

Perturbations around co-homogeneity 1 backgrounds in the variables
$x$ may be dimensionally reduced in a straight-forward way. The
tensor indices of each field are decomposed into $x$ and the rest.
For instance, the metric perturbation tensor $h_{\mu\nu}$ decomposes
into the $x$-tensor $h_{xx}$, the $x$-vectors $h_{xi}$ and
$x$-scalars $h_{ij}$ where $i$ runs over the other coordinates. The
dependence of the fields on the homogeneous coordinates is expanded
in harmonic functions, substituted back into the action and the
homogeneous coordinates are integrated over. Finally the gauge
functions are also dimensionally reduced via reduction of indices
and harmonic expansion.

Continuous space-time isometries of the background become local
(gauge) symmetries of the action while discrete isometries (which
operate on the boundary) become global symmetries. In a gauge
theory, on the other hand, all the symmetries of the background
become global symmetries. Accordingly, the perturbations are charged
under the isometries which they break. Since the quadratic action
may include only interactions between fields in the same
representation (more precisely, in a pair of complex conjugate
representations), we recognize an important mechanism that decouples
the action into sectors.

We shall now apply these principles to two main cases of physical
importance.

\subsection{Spherical symmetry; Cosmology}

\noindent {\bf Spherical symmetry}. Here we shall describe the
counting of the dynamic degrees of freedom for the 4d \Schw black
hole, in the terms of this paper, and postpone the detailed analysis
and the rest of the results to a later publication (work in
progress).

Working in standard $r,t,\Omega, ~\Omega \equiv (\theta,\varphi)$
\Schw coordinates the isometries are $U(1)_t \times SO(3)_\Omega$.
The essential dimension of the background which we called $x$ in the
general case is $r$, and the homogeneous 3 space fibered over $r$ is
$\IR_t \times \IS^2_\Omega$.

We wish to dimensionally reduce the action to a 1d action in $r$.
The metric perturbation tensor $h_{\mu\nu}$ is decomposed into
$h_{rr},\, h_{rt},\, h_{tt},\, h_{ri},\, h_{ti},\, h_{ij}$ where
$i,j$ run over $\theta,\varphi$. A harmonic decomposition of the
fields means a Fourier expansion in the $t$ direction and spherical
harmonic decomposition in the angular coordinates, thus transforming
each field $h(r,t,\Omega) \to \sum_{\omega,l m}\, h_{\omega,l
m}(r)$. The gauge functions split as well $\xi_\mu \to (\xi_r,\,
\xi_t,\, \xi_i)$.

An important role is played by the $\IZ_2$ parity isometry. Fields
whose parity is $(-)^l$ are called ``even'' while those with
$(-)^{l+1}$ are called ``odd''. Accordingly the action decouples
into an even part and an odd part. Table \ref{table1} summarizes the
counting of fields and demonstrates the formula for $\dim D$ (item D
of theorem \ref{thm}).

\begin{table}[t!]
\centering $\begin{array}{l||c|c}
  {\rm Sector}   &  {\rm ``Even''}  & {\rm  ``Odd''} \\
 \hline \hline
 {\rm fields} & h_{rr}~ h_{rt},~ h_{tt} &             \\
            & h_{ri}^{(1)}~ h_{ti}^{(1)}~ {\rm tr}(h_{ij})~ h_{ij}^{(1)} & h_{ri}^{(2)}~
            h_{ti}^{(2)}~ h_{ij}^{(2)} \\
 \mbox{gauge functions} & \xi_r~ \xi_t~ \xi_i^{(1)} & \xi_i^{(2)} \\
 \hline
  n_F  &  7 & 3 \\
  n_G &  3 & 1  \\
  \hline
  \dim(D)= n_F-2\, n_G  &  1 & 1  \\
  \hline
  DG , ~~~|DG|=n_G  & h_{rr}~ h_{rt}~ h_{ri}^{(1)} & h_{ri}^{(2)}
\end{array}$
\caption[]{4d \Schw space-time: counting the perturbation fields as
an application of the general procedure of section
\ref{theorem-section}. The action decouples into ``even'' and
``odd'' sectors according to the charge under the parity isometry.
We list the fields and gauge functions in each sector. Altogether we
``predict'' 1 dynamical even field (the field appearing in the
Zerilli master equation) and 1 dynamical odd field (the field
appearing in the Regge-Wheeler master equation). Fields which are
vectors with respect to the sphere (those which carry an $i$ index)
are decomposed into two families of spherical harmonics: one which
is a gradient and another which is divergence-free, accordingly the
fields are denoted by a $(1)$ or $(2)$ superscript. A similar
notation holds for the tensor field $h_{ij}$. Finally we list the
derivatively-gauged (and hence mutually algebraic) fields in each
sector in the $DG$ row.} \label{table1}
\end{table}

\presub {\bf Cosmology}. In homogeneous cosmology the essential
coordinate is $t$, the homogenous fiber is $\IR^3$ with its 3d
Poincar\'{e} isometry. The 3d momentum operator may be diagonalized
on the perturbation space, yielding the wave-number 3-vector
$\vec{k}$. $\vec{k}$ breaks the 3d Poincar\'{e} group down to the 2d
Poincar\'{e} group. The translation part of the latter is not
broken, but the $SO(2)$ rotations may be broken according to the
$SO(2)$ tensor type of the perturbation. Accordingly we distinguish
three sectors which the action decouples into: scalar, vector and
tensor. Note that while in the standard presentation the terms
scalar, vector and tensor refer to $SO(3)$ here they refer to the
``little group'' $SO(2)$. Table \ref{table2} summarizes the counting
of fields and demonstrates the formula for $\dim D$ (item D of
theorem \ref{thm}).

\begin{table}[t!]
\centering $\begin{array}{l||c|c|c}
  {\rm Sector}   &  {\rm scalar}  & {\rm vector} & {\rm tensor} \\
 \hline \hline
 {\rm fields} & h_{rr}~ h_{rz},~ h_{zz} ~{\rm tr}(h_{ij}) &  h_{ri} ~ h_{zi} & h_{ij}          \\
 \mbox{gauge functions} & \xi_r~ \xi_z  & \xi_i  & \mbox{---} \\
 \hline
  n_F  &  4 & 2 * 2  & 1*2 \\
  n_G &  2 & 1*2 & \mbox{---}  \\
  \hline
  \dim(D)= n_F-2\, n_G  &  0 & 0 & 2  \\
  \hline
  DG, ~~~|DG|=n_G & h_{tt}~ h_{zt} & h_{ti} & \mbox{---}
\end{array}$
\caption[]{Homogenous cosmology: counting perturbation fields as an
application of the general procedure of section
\ref{theorem-section}. The action decouples into scalar, vector and
tensor sectors according to the charge under the $SO(2)$ ``little
group'' isometry. We list the fields and gauge functions in each
sector, remembering in 4d to associate 2 polarizations with each
$SO(2)$ vector or tensor. Altogether we ``predict'' 2 dynamical
tensor fields (the 2 polarizations of the graviton). If the
$\vec{k}$ vector is taken to lie in the $z$ direction and $i,j$ run
over $x,y$. Finally we list the derivatively-gauged (and hence
mutually algebraic) fields in each sector in the $DG$ row. If we
couple gravity to matter, then as long as the energy-momentum tensor
of the matter is spatially isotropic, the matter is guaranteed to
couple only to the 2 algebraic scalar modes.} \label{table2}
\end{table}

\newpage
\noindent {\bf Acknowledgements}

I would like to thank Vadim Asnin and Michael Smolkin for
collaboration on related work.

This research is supported in part by The Israel Science Foundation
grant no 607/05 and by the Binational Science Foundation
BSF-2004117.

\end{document}